\documentclass[fleqn,10pt]{wlscirep}
\usepackage[utf8]{inputenc}
\usepackage[T1]{fontenc}
\usepackage{multirow}
\usepackage{listings}
\title{Accessible Data Curation and Analytics for International-Scale Citizen Science Datasets}
\author [1, *]{Benjamin Murray}
\author [1]{Eric Kerfoot}
\author [1]{Mark S. Graham}
\author [2]{Carole H. Sudre}
\author [1]{Erika Molteni}
\author [1]{Liane S. Canas}
\author [1]{Michela Antonelli}
\author [1]{Kerstin Klaser}
\author [3]{Alessia Visconti}
\author [4]{Andrew T. Chan}
\author [5]{Paul W. Franks}
\author [6]{Richard Davies}
\author [6]{Jonathan Wolf}
\author [3]{Tim Spector}
\author [3]{Claire J. Steves}
\author [1, $\dag$]{Marc Modat}
\author [1, $\dag$]{Sebastien Ourselin}
\affil[1]{King's College London, School of Biomedical Engineering \& Imaging Sciences, London, SE1 7EU, United Kingdom}
\affil[2]{University College London, MRC Unit for Lifelong Health and Ageing, Department of Population Health Sciences, London, WC1E 7HB, United Kingdom}
\affil[3]{King's College London, Department of Twin Research and Genetic Epidemiology, Westminster Bridge Road, London, SE1 7EH, United Kingdom}
\affil[4]{Massachusetts General Hospital, 55 Fruit Street, GRJ 825C, Boston, MA 02116, United States}
\affil[5]{Lund University, Diabetes Centre, CRC, SUS Malmö, Jan Waldenströms gata 35, House 91:12. SE-214 28 Malmö, Sweden}
\affil[6]{Zoe Global Limited, 164 Westminster Bridge Road, London, SE1 7RW, United Kingdom}

\affil[*]{corresponding author(s): Benjamin Murray (benjamin.murray@kcl.ac.uk)}

\affil[$\dag$]{Equal contribution}

\begin{abstract}
The Covid Symptom Study, a smartphone-based surveillance study on COVID-19 symptoms in the population, is an exemplar of big data citizen science. Over 4.7 million participants and 189 million unique assessments have been logged since its introduction in March 2020. The success of the Covid Symptom Study creates technical challenges around effective data curation for two reasons. Firstly, the scale of the dataset means that it can no longer be easily processed using standard software on commodity hardware. Secondly, the size of the research group means that replicability and consistency of key analytics used across multiple publications becomes an issue. We present ExeTera, an open source data curation software designed to address scalability challenges and to enable reproducible research across an international research group for datasets such as the Covid Symptom Study dataset. 
\end{abstract}
\begin{document}

\flushbottom
\maketitle

\thispagestyle{empty}

\section*{Introduction}

Mobile applications have enabled citizen science ~\cite{silvertown2009new, newman2012future, follett2015analysis, Heigl8089} projects that can collect data from millions of individuals. The Covid Symptom Study ~\cite{Drew1362}, a smartphone-based surveillance study on self-reported COVID-19 symptoms started in March 2020, is an exemplar of citizen science. As of 14th October 2020, the study contains 189 million self-assessments collected from more than 4.7 million individuals, provided as daily comma separated value (CSV) snapshots, that are made available to both academic and non-academic researchers to facilitate COVID-19 research by the wider community.

The primary challenge in curating and analysing the data is its scale. Scale adds complexity to otherwise simple operations. Python-based scientific computing libraries such as Numpy ~\cite{walt2011numpy, harris2020array} and Pandas ~\cite{mckinney-proc-scipy-2010} are ubiquitous in the academic community, but they are not designed to scale to datasets larger than the amount of Random Access Memory (RAM) on a given machine. Commodity hardware, as of 2020, is typically equipped with 16 to 32 GB of memory. Although it is possible to procure hardware with more RAM, this involves server-grade pricing above 64 GB of memory, and doubling memory only doubles the size of the data that can be handled.

Alternatively, data can be moved to a datastore; either traditional, relational databases such as PostgreSQL or other types of datastore including key-value stores such as CouchBASE. Each type of datastore comes with its own design philosophy and set of performance trade-offs ~\cite{stonebraker2010sql}. Datastores are capable of running queries on larger-than-RAM datasets, but all come with the need to learn either a new language or application programming interface (API), and an installation and maintenance burden, whether installed locally or deployed in a cloud.

% Although most of the principal Python libraries that support data science are not designed to work with datasets that are significantly larger than RAM, they provide a rich set of functionality that can be built on top of in order to provide key algorithms that are critical for scaling to drive-scale datasets, whilst still leveraging the existing Python ecosystem. The resulting software can be given an API that is familiar to users of the Python ecosystem.

% Although most of the principal Python libraries that support data science are not designed to work with datasets that are significantly larger than RAM, they provide a rich set of functionality that can be built on top of, as well as demonstrating design choices that have proven successful with the community. By focusing on the provision of key algorithms that are critical for scaling beyond RAM, whilst still leveraging the existing Python ecosystem it is possible to create a highly scalable data analysis software without requiring a large, dedicated development team. The resulting software can be given an API that is familiar to users of the Python ecosystem.

Although most of the principal Python libraries that support data science are not designed to work with datasets that are significantly larger than RAM, they provide a rich set of functionality that can be built on top of, as well as demonstrating design choices that have proven successful with the community. By focusing on the provision of key algorithms that are critical for scaling beyond RAM, whilst still leveraging the existing Python ecosystem it is possible to create a highly scalable data analysis software with an API that is familiar to users of the Python ecosystem.

Aside from scale, data curation includes other significant engineering challenges.

Raw datasets are typically noisy and may contain conflicting and inconsistent values. Erroneous values, changing schemas and multiple contemporary app versions all add complexity to the task of cleaning and consolidating datasets for downstream analysis.

Another major challenge is reproducibility of analyses, especially across a large research team. When data cleaning and generation of analytics is done in an ad-hoc fashion, it is easy to generate subtly different derivations of the same measure, causing difficulties in reconciling research efforts across multiple groups and research outputs. Full reproducibility requires algorithms to be treated as immutable, so that the application of a particular algorithm to a particular snapshot of data guarantees identical results, hardware notwithstanding.

Public datasets are often delivered as a series of timestamped snapshots. The snapshots are typically memory-less; if two snapshots have corresponding entries that are modified over time, one can only determine the delta by comparing the snapshots, which exacerbates scaling issues. Updating an analysis from one snapshot to another without such comparison compromises the interpretability of the updated results in a way that is not visible when working with individual snapshots.

% To address the above challenges, we have created ExeTera: a prototype software that enables sophisticated, auditable, and reproducible data curation and analytics by distributed research teams on international-scale datasets. ExeTera is implemented with a Python API designed to be familiar to data analysts and researchers who use Python’s scientific computing tools, while enabling the processing of datasets approaching terabyte scale.

% ExeTera has initially been implemented to provide data curation and analysics for the Covid Symptom Study. In this work, we evaluate ExeTera's performance on this dataset and demonstrate that it is suitable for analysis of tabular datasets approaching terabyte scale in the general case.

To address the above challenges, we have created ExeTera; a software that enables sophisticated, auditable, and reproducible data curation and analytics by distributed research teams, through a Python API designed to be familiar to data analysts and researchers who use Python's scientific computing ecosystem. ExeTera has initially been implemented to provide data curation and analytics for the Covid Symptom Study. In this work, we evaluate ExeTera's performance on this dataset and demonstrate that it is suitable for analysis of tabular datasets approaching terabyte scale in the general case.

\section*{Results}

ExeTera's performance has been benchmarked against a combination of artificial data and data from the Covid Symptom Study. We examine performance and scalability of ExeTera for key operations relative to Pandas, the most popular Python library for analysing tabular data. We also demonstrate Exetera's ability to generate a journaled dataset from snapshots allowing longitudinal analysis that is able to account for destructive changes between corresponding rows of the different snapshots. Finally, we present an example of ExeTera's analytic capabilities.

Performance was measured on an AMD Ryzen Threadripper 3960x 24-core processor with 256 GB of memory. All processes were limited to 32GB of memory. The data was read from and written to a 1TB Corsair Force MP600, M.2 (2280) PCIe 4.0 NVMe SSD.

\subsection*{Importing to ExeTera}

ExeTera has an import function that is used to convert a set of source CSV files to HDF5 ~\cite{hdf5}, given a JSON (JavaScript Object Notation) schema document describing the datatypes. This operation streams the CSV files, transforming the string source field data into one or more target fields in the HDF5 dataset. Table \ref{tab:import} shows the time taken and resulting table row counts for importing the Covid Symptom Study snapshot from the 8th October 2020.

\subsection*{Comparison of ExeTera and Pandas for key operations}

We compare ExeTera with Pandas for two key performance and scalability tests; the ability to load subsets of a dataset and the ability to perform joins across different tables in a dataset.

\subsubsection*{Data subset reading}

In the general case, a typical piece of analysis will only involve a fraction of the available fields in a dataset. We test the ability of ExeTera and Pandas to load subsets of the Covid Symptom Study dataset, from the patient and assessment tables respectively.

\paragraph{Covid Symptom Study Patient Table}

To generate the results in Table \ref{tab:read_patent_fields}, we load 1, 2, 4, 8, 16, and 32 fields respectively from the patient table of the Covid Symptom Study dataset, containing approximately 4.7 million entries. Table \ref{tab:read_patent_fields} shows that ExeTera (given its use of HDF5 as its serialized format) is orders of magnitude faster than CSV due to each entire field being loadable with effectively zero parsing overhead and as monolithic reads. CSV on the other hand, must be parsed in its entirety and the resulting collections incrementally built. CSV files are loaded through Panda’s \verb|read_csv| function.

\paragraph*{Covid Symptom Study Assessment Table}

To generate the results in Table \ref{tab:read_assessment_fields}, we load 1, 2, 4, 8, 16, and 32 fields respectively from the assessment table of the Covid Symptom Study dataset, containing approximately 189 million entries. The results in Table \ref{tab:read_assessment_fields} show that Pandas is unable to load multiple fields at this row count. Again, CSV files are loaded through Panda’s \verb|read_csv| function.

\subsubsection*{Left join operation}

The join operation is the most memory-intensive operation that is typically carried out on tabular datasets. ExeTera implements highly scalable versions of left, right and inner joins; left join is the most typical join performed so we focus on its performance here.

\begin{figure}[ht]
    \centering
    \includegraphics[width=\linewidth]{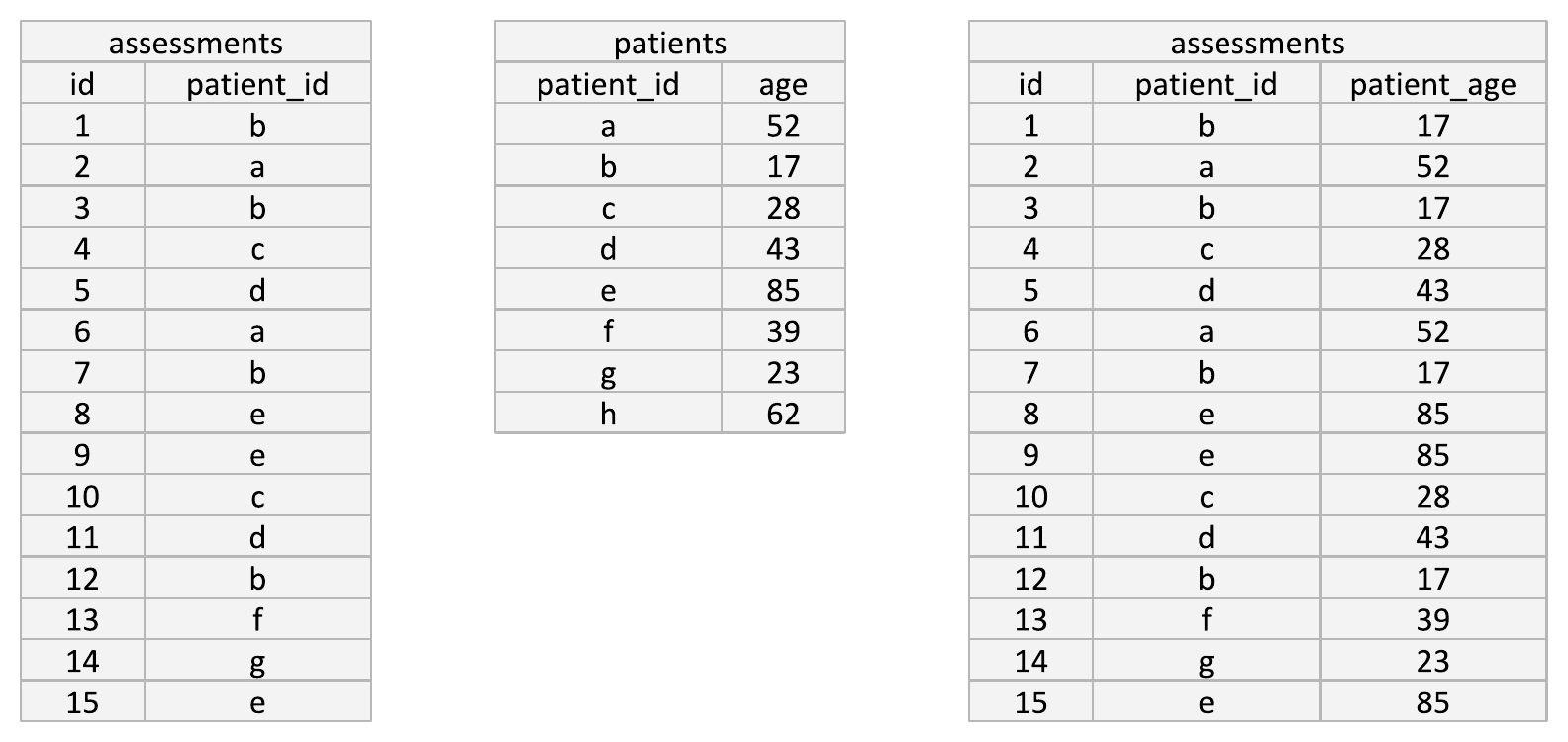}
    \caption{A left join of a simplified dummy patent and assessment dataset. The left join is from patients on the right to assessments on the left to generate a \texttt{patient\_age} field in assessment space.}
    \label{fig:left_join_p_to_a}
\end{figure}

A left join operation involves the mapping of data from one table onto another table, based on the key relationships between the two tables. Values in the right table are mapped to values in the left table. For example, patient data may be joined to assessment data so that assessments can be processed with patient-level features such as age or BMI (body mass index). An illustrated example of a left join can be seen in Figure \ref{fig:left_join_p_to_a}.

To test the performance of the join operator when ExeTera and Pandas are used, we generate a dataset composed of a left primary key (int64), a right foreign key (int64) and 1, 2, 4, 8, 16, and 32 fields respectively of random numbers corresponding to entries in the right table (int32). The first ten rows of the join dataset in the two field case is shown in Table \ref{tab:join_data_example}. The resulting join performance for both is shown in Table \ref{tab:left_join}. The left and right tables have the same number of rows for each test, listed in the first column.

It should be noted that ExeTera organises data into groups, which may be considered logical tables, but the user typically loads the data on a field by field basis when working with it. ExeTera’s merging API allows the user to think in a tabular fashion by accepting tuples of ExeTera fields, which only load their contents when specifically requested. ExeTera achieves its scalability through two techniques; firstly, by calculating the map and then applying it to each field in turn, and secondly, by exploiting the natural sorted order to convert the mapping into a streaming merge. The result is a merge operation that is fast and scales to billions of rows.

\subsection*{Journaling operation}
ExeTera can combine snapshots of datasets to create a journaled dataset, keeping multiple, timestamped copies of otherwise destructive changes to corresponding records between the snapshots. Table \ref{tab:journaling} shows the results of journaling together shapshots of the Covid Symptom Study from the August 1st 2020 and September 1st 2020, and the time taken to do so.

\subsection*{Analytics}
\begin{figure}[ht]
    \centering
    \includegraphics[width=\linewidth]{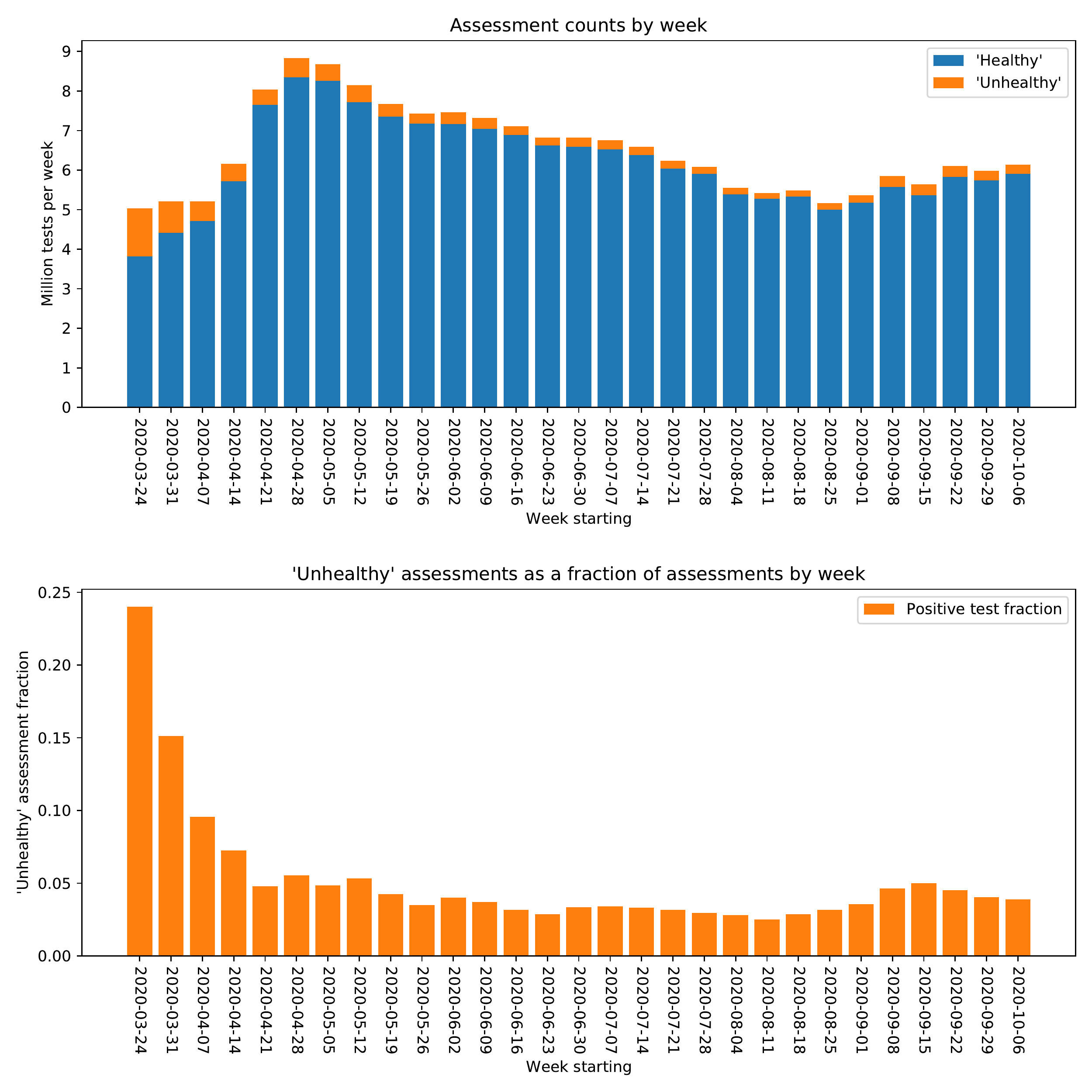}
    \caption{Seven day summary of assessments from the Covid Symptom Study snapshot dated 14th October 2020. The upper chart shows the number of assessments, coloured by whether the patient logged as healthy or unhealthy. The lower chart shows the assessments logged as unhealthy as a fraction of assessments logged for that seven day period}
    \label{fig:assessments_summary}
\end{figure}

ExeTera provides the ability to load data very efficiently, as seen in Table \ref{tab:read_patent_fields} and \ref{tab:read_assessment_fields}. Once loaded, analytics can be performed through use of libraries such as Numpy and Matplotlib ~\cite{Hunter:2007}, using tools that researchers are familiar with, such as Jupyter Notebook~\cite{Kluyver:2016aa}. Figure \ref{fig:assessments_summary} shows a histogram of healthy and unhealthy assessment logs bucketed into seven day periods that must parse 189 million assessments to generate its results.

\section*{Discussion}

In this work, we present ExeTera, a data curation and analytics tool designed to provide users with a low complexity solution for working on datasets approaching terabyte scale, such as national / international-scale citizen science datasets like the Covid Symptom Study. ExeTera makes this possible without the additional complexity of server-based datastores, and thus simplifies access to such datasets for both academic and independent researchers who are versed in the Python scientific programming ecosystem.

ExeTera provides features for cleaning, journaling, and generation of reproducible processing and analytics, enabling large research teams to work with consistent measures and analyses that can be reliably recreated from the base data snapshots. Its ability to store multiple snapshots in a journaled format enables researchers to perform full longitudinal analysis on otherwise unjournaled datasets and facilitates the ability to move between snapshots whilst being able to properly explore the impact of doing so on analyses.

ExeTera has been a key part in enabling analysis of the Covid Symptom Study dataset, including being used for analysis in the following manuscripts ~\cite{Ni_Lochlain_et_al, Costeira_et_al, Bowyer_et_al, Bataille_et_al}.

% ExeTera makes this possible without the additional complexity of server-based datastores, and thus enables access to such datasets for both academic and independent researchers. It  allows the handling of national-scale citizen science datasets without having to leave programming paradigms that are familiar to Python-based researchers, through an API that is familiar to researchers versed in Numpy and Pandas.

% Although ExeTera was developed to provide data curation for researchers working on the Zoe Symptom Study, this software is being developed to be generally applicable to large-scale relational datasets for researchers who work in Python. However, this requires further work to separate core ExeTera functionality from Covid Symptom Study-specific functionality. While our goal is to closely emulate a Pandas-style API, additional work is still required to ensure this is available to the user across the whole API. The sorting and journaling APIs, for example, are accessed through the low-level operations API rather than through abstractions to more closely emulate Pandas, and work is ongoing to make the API more consistent in the presentation and manipulation of the data as logical tables.

Although ExeTera was developed to provide data curation for researchers working on the Covid Symptom Study, this software is being developed to be generally applicable to large-scale relational datasets for researchers who work in Python. As such, ExeTera core functionality is provided as a python package that is imported by ExeTeraCovid, the collection of algorithms, scripts, and other resources specific to the Covid Symptom Study.

While our goal is to closely emulate a Pandas-style API, additional work is still required to ensure this is available to the user across the whole API. The highly scalable sorting API and journaling API, for example, are accessed through the low-level operations API rather than through abstractions to more closely emulate Pandas, and work is ongoing to make the API more consistent in the presentation and manipulation of the data as logical tables.

ExeTera is currently built on top of Numpy and Pandas functionality. The ability to scale is provided through implementations of key operations that can stream large collections from drive. This enables processing of large datasets, but, at present, ExeTera doesn’t take advantage of  multiple cores or processors, nor is it able to process across a cluster. Future development will make use of Dask ~\cite{dask} a library designed to convert operations on Numpy and Pandas to a directed acyclic graph of sub-operations that can be distributed to multiple cores and nodes, and focus on provision of the elements of scalable processing, such as multi-key argsorts and tableless merges that Dask lacks. Dask is also integrated with more specialised back-ends such as Nvidia’s RAPIDs ~\cite{rapids}, which enables execution of distributed graph processing across GPU clusters. The integration of Dask should allow ExeTera to execute over datasets well into the multi-terabyte range.

Reproducibility relies on immutability of the algorithms deployed during analysis. At present, ExeTera achieves this through a convention that algorithms are treated as immutable once implemented and deployed. A more robust system is being designed to provide algorithmic immutability without this constraint.

HDF5 has proven to be fragile to interruptions while data is being written to it; an interrupted write is capable of rendering the entire dataset unreadable and so all writes must be protected from user interrupts and other exceptions. Additionally, HDF5 does not allow the space from deleted fields to be properly reclaimed from the dataset. This must be done separately through use of the ‘h5repack’ tool. The primary benefit of HDF5 to ExeTera is its ability to flexibly store contiguous fields of data for rapid reading. This can also be achieved through use of alternative columnar data formats such as ORC ~\cite{Floratou2018} or Parquet ~\cite{vohra2016, Floratou2018} or, alternatively, through use of the file system as a datastore. Storing individual fields as serialized Numpy arrays and field metadata in JSON allows for transparent, robust dataset serialization that can be explored on the file system.

% To conclude, ExeTera enables reproducible generation of results from timestamped snapshots and explicitly versioned algorithms. It provides the ability to build and query datasets constructed from a series of snapshots over time, allowing for longitudinal analyses that can take account of otherwise destructive changes to the dataset.

ExeTera is made available as open source software released under Apache License 2.0.
% ExeTera is a critical part of the larger Covid Symptom Study research group’s ability to analyse the dataset and publish papers and is currently under active development and 

\section*{Methods}

\subsection*{The Covid Symptom Study dataset}

% Data were collected using the COVID Symptom Study app, developed by Zoe Global Ltd with input from King’s College London, the Massachusetts General Hospital,​ Lund University, Sweden and Uppsala University, Sweden

ExeTera has been developed by King’s College London (KCL) to provide data curation for the Covid Symptom Study dataset. The dataset is collected using the Covid Symptom Study app, developed by Zoe Global Ltd with input from King's College London, the Massachusetts General Hospital, Lund University Sweden, and Uppsala University, Sweden. It is a response to the COVID-19 pandemic based on epidemiological surveillance via smartphone-based self-reporting. It asks citizens from the UK, US, and Sweden to use a mobile application to log symptoms, record COVID-19 test results and answer lifestyle and occupational questions. The Covid Symptom Study dataset has generated insights into COVID-19 that have gone on to inform governmental policies for handling of the disease ~\cite{menni2020real, Drew1362, nguyen2020risk, zazzara2020delirium}. In the UK, the App Ethics has been approved by KCL ethics Committee REMAS ID 18210, review reference LRS-19/20-18210 and all subscribers provided consent. In Sweden, ethics approval for the study is provided by the central ethics committee (DNR 2020-01803).

As of the 14th October, 2020, the dataset is composed of four tables:

\textbf{Patients:} 4.701 million patients with 224 data fields. Patient records store data such as the patients' physiological statistics, long-term illnesses, lifestyle factors, location and other data that only occasionally changes, at the patient level.

\textbf{Assessments:} 189.2 million assessments with 66 fields. Patients are asked to give regular assessments through the app that cover their current health status and symptoms, aspects of their lifestyle such as potential exposure to COVID-19, and, in early versions of the schema, any COVID-19 tests that they have had.

\textbf{Tests:} 1.455 million tests, with 26 fields. Test records are kept for each COVID-19 test that a patient has had along with the evolving status of that test (typically ‘waiting’ to some result).

\textbf{Diet:} 1.563 million diet study questionnaires with 104 fields. These ask people at several time points about their dietary and lifestyle habits.

Assessments, tests and diet study questionnaires are mapped to patients via ids that serve as foreign keys.

This dataset is delivered as daily snapshots in CSV format. As of 14th October 2020, the daily snapshot is 42.2 GB in size, and the accumulated daily snapshots are over 3TB in size. The dataset, excepting fine-grained geolocation data, is publicly available at \hyperlink{https://healthdatagateway.org}{https://healthdatagateway.org}.

\subsection*{Scalability as prerequisite}
In order to successfully curate the Covid Symptom Study data, it is necessary to be able to handle data that cannot fit into RAM. Data size and structure, and the set of operations needed to handle the dataset have to be addressed. We can define three scale domains that necessitate a change of approach.

\subsubsection*{RAM Scale (1GB to 16 GB)}
This is the scale at which the dataset entirely fits in the computer’s RAM. Commodity laptops and desktops used by researchers typically have between 16 and 32 GB of RAM. Loading the data can inflate its memory footprint depending on the datatypes used, and operations can multiply memory requirements by a small constant factor, but provided peak memory usage does not dramatically exceed RAM, researchers can make use of programming languages with numerical / scientific libraries such as Numpy or Pandas to effectively analyse the data.

\subsubsection*{Drive Scale (16 GB to 1 TB)}
At drive scale, only a portion of the dataset can fit into RAM at a given time, so specific solutions are required to effectively stream the dataset from drive to memory. Datastores become a more compelling option at this scale, as they already have memory efficient, streaming versions of the operations that they support, but their usage may not be desirable due to the need to learn a new language or API, and the installation and maintenance burden they represent. This is the scale of dataset that ExeTera currently targets.

\subsubsection*{Distributed Scale (> 1 TB )}
At distributed scale, the use of server-based datastores is typically mandatory. It becomes necessary to redesign operations to exploit distributed computing across many nodes. Selection of appropriate datastore technology becomes critical, with specific datastore technologies addressing different roles within the overall system. This scale will be targeted by ExeTera in future development.

\subsection*{A model data curation pipeline}
The ExeTera software provides functionality that enables a data curation pipeline incorporating data curation best practice. The pipeline has the following steps:
\begin{itemize}
  \item Import / preliminary data cleaning and filtering
  \item Journaling of snapshots into a consolidated dataset
  \item Generation of derived data and analytics 
\end{itemize}

The first two stages are generic operations that apply to any tabular dataset being imported into ExeTera. The third stage is specific to a given dataset, such as the Covid Symptom Study.

\subsubsection*{Import / Preliminary Cleaning and Filtering}
The import process converts CSV data to a binary, columnar format, discussed further in the ‘Implementation’ subsection, that is many orders of magnitude more time efficient for querying in most cases. As part of this process, the data is converted from strings to data types defined by a JSON schema file.

The mapping of data types is from a single CSV field to one or more strongly typed fields. How this is done is determined by the JSON schema and the type specified in the schema.

\paragraph*{Fixed string fields}

Fixed string fields contain string data where each entry is guaranteed to be no longer than the length specified by the field. Fixed string fields can handle UTF8 unicode data, but this is encoded into bytes and so the specified length must take into account the encoding of the string to a byte array.

\paragraph*{Indexed string fields}

Indexed string fields are used for string data where the strings may be of arbitrary length. The data is stored as two arrays; a byte string of all of the strings concatenated together, and an array of indices indicating the offset to each entry.

\paragraph*{Numeric / logical fields}

Fields which contain a combination of strings to be converted to strongly typed values and empty values, for example “”, “False”, “True” are converted to the appropriate numeric / logical dataset and a corresponding filter field indicating whether a value is present for a given row of the field. Fields containing string values can be processed as categorical fields if specified as such.

\paragraph*{Categorical fields}

Categorical fields map a limited set of string values to a corresponding numeric value. A key is stored along with the field providing a mapping between string and number. A categorical field can also be specified as leaky, in the case that the field is a mixture of categorical values and free text. In this case, a value is reserved to indicate that a given row doesn’t correspond to a category, and an indexed string field is created for free text entries.

\paragraph*{Datetime / date fields}

Datetime fields store date times as posix timestamps in double precision floating point format. The schema can also specify the generation of a ‘day’ field quantising the timestamp to the nearest day, and can also specify whether the field contains empty values, in which case as filter is also generated, as with numeric fields.

% new for re-submission 1
\paragraph{Schema file format}
Importing data from CSV requires a schema file that describes the fields and the type conversions that they should undergo. The ExeTera schema file format is a JSON format. Each table is described by entry inside of a JSON dictionary labeled \texttt{schema}. Each entry in this dictionary is the name of the table followed by the table descriptor. This has up to three entries. The first is \texttt{primary\_keys}, which lists zero or more fields for the dataset that together represent the primary key for the table. The second is \texttt{fields} and contains all the field descriptors for the table. The third is \texttt{foreign\_keys} and contains the names of foreign keys in the table and which other tables they relate to.

% new for re-submission 2
\paragraph {Schema file field entries}
The schema file entries themselves contain at minimum a \texttt{field\_type} entry, and depending on the specific field type, require additional entries. Figure \ref{fig:schema_file_example} shows an illustrative, minimal example of a schema file. A full specification can be found in the ExeTera github wiki, as of the time of writing.

\subsubsection*{Journaling of snapshots into a consolidated dataset}

Data for the Covid Symptom Study project is delivered as a series of timestamped snapshots. The unanonymised data generated by the Covid Symptom Study app is stored in a relational database or similar datastore, that is not accessible to query by the broader research community. Instead, the data needs to first be anonymised and bulk exported to CSV format. The database is a live view of the dataset, however; users can update data through the app, and, unless the database is explicitly journaled and each entry made immutable, the prior states are erased. As such, a row corresponding to a given entity in two different snapshots can be contain conflicting values.

\begin{figure}[ht]
    \centering
    \includegraphics[width=\linewidth]{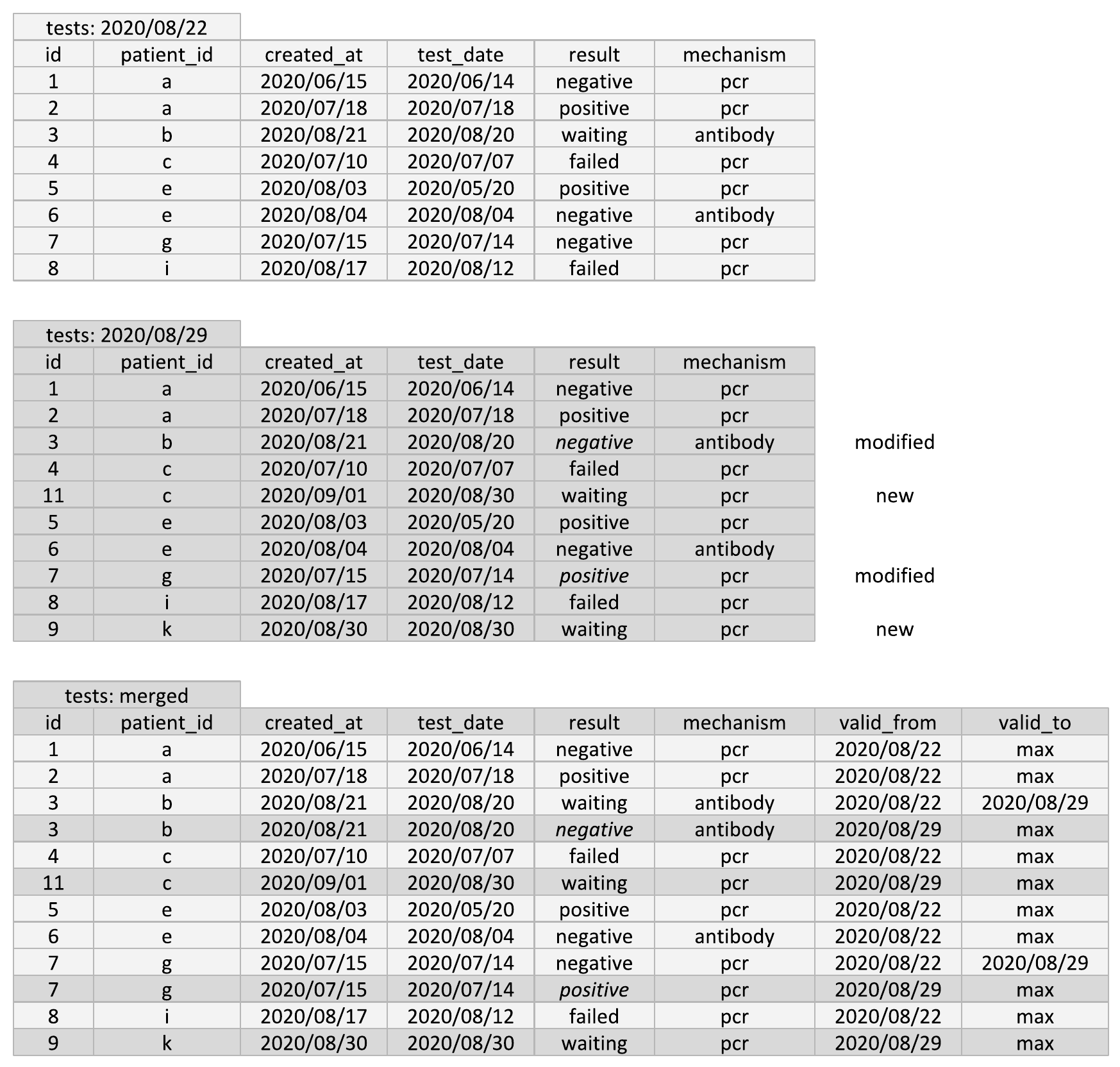}
    \caption{Construction of a journaled dataset Two snapshots of a simplified dummy dataset representing COVID-19 tests, one from 2020/08/22 and one from 2020/08/29 are used to construct a journaled dataset.}
    \label{fig:test_journaling}
\end{figure}

When each snapshot is large, the scaling problem is exacerbated by having to reconcile multiple snapshots. The Covid Symptom Study dataset does not have a field that reliably indicates whether the contents of a given row have changed and so determining whether a row at time $t$ has changed relative to a row at time $t+1$ requires a full comparison of all common fields. An example of this can be seen in Figure \ref{fig:test_journaling}. 

\subsubsection*{Generation of derived data and analytics}

In addition to the initial data cleaning performed during import, it is useful to perform application specific cleaning and generate ancillary fields that are widely used for downstream analyses. This helps to ensure consistency across analytics and reduces scripting complexity for new users.

\subsection*{Covid Symptom Study-Specific Cleaning and Processing}

The Covid Symptom Study data schema has seen rapid iteration since its inception, due to a number of factors. Firstly, the initial app was rapidly released to allow users to contribute as soon as possible after the pandemic was declared. Secondly, the evolving nature of the pandemic, particularly around prevalence in the population and availability and type of tests has necessitated structural changes to the schema. Thirdly, this dataset is novel in terms of its scale and deployment for Epidemiological analysis, and the initial wave of papers published by the research group has fed back into the schema.

Public health surveillance campaigns such as the Covid Symptom Study impose time constraints to software development, with frequent changes in database structure and intense versioning to accommodate iterative refinements. The evolving epidemiology of COVID-19, the response of governments and populations to the pandemic, and academic responses to papers based on the dataset all shape the questions that are added to or removed from the app over time.

The dataset is only minimally validated at source. The fields often contain data of mixed type, and different fields can be in mutual contradiction. Numeric values are only validated for type rather than sensible value ranges. Furthermore, the dataset contains multiple competing schema for the same underlying data, and the app version is tied to the schema version, so users who are using older versions of the app are still contributing to otherwise retired schema elements. As such, a considerable amount of data cleaning and processing is required in order to extract data suitable for analysis.

\subsubsection*{Schema changes}

The handling of COVID-19 tests in the dataset is an example of the complexity created by changes to the schema. Testing was initially reported as an assessment logging activity, but this came with a number of problems. Firstly, a test needed to be logged on the day it was taken for the assessment date to be treatable as the test date. Secondly, some users interpreted the test field as something to be logged only when they took a test or received the result, whilst other users filled in intermediate assessments with the pending status. Thirdly, this system did not allow for users to enter multiple tests unambiguously. Whilst this was not a problem in the initial months of the pandemic, the ramping up of test availability necessitated a solution.

A new test table was introduced in June 2020, giving each test a unique id to allow multiple tests for each patient. However, existing tests recorded in the old schema were not connected with new test entries, although many users re-entered old test results in the new test format. Furthermore, new tests continued to be added by users in the old, assessment-based schema format, logging on previous versions of the app. As such, there is no unambiguous way of determining whether tests in the old format are replicated by tests in the new format. This is an example of a postprocessing activity with no unambiguously correct output, which therefore requires at least a single, agreed upon algorithm to be consistently deployed to avoid inconsistencies between related analyses.

\subsubsection*{Validation of user-entered values (weight, height, BMI)}

In the Covid Symptom Study app, user-entered numeric values are only validated to ensure that they are numeric, as of the time of writing. There are no validations of sensible ranges given the user-selected units of measurement. Some users enter incorrect values, and some users enter values that appear sensible but only in some other unit (1.8 is a plausible height if the user is entering height in metres, for example).

\subsubsection*{Quality metrics for test mechanism}

The covid test table has a ‘mechanism’ field where the user is free to either select a categorical value indicating the test mechanism, or enter free text relating to the test mechanism. Some free text clearly indicates the test type, whereas other free text entries only infer the test type weakly, through inference such as ‘home test kit’.  As such, a set of gradated flags are generated that indicate the quality of the categorisation.

\subsubsection*{Generation of daily assessments}

In case of multiple daily entries by the users, these assessments can optionally be quantised into a single daily assessment that, for symptoms, corresponds to the maximum value for each symptom that the user reported in the day. This considerably simplifies many downstream analyses.

\subsubsection*{Generation of patient-level assessment and test metrics}

Analysis often involves the filtering of patients that are categorised by aspects of the assessments and tests that they have logged. These include metrics such as whether the patient logged as being initially healthy, or whether they have ever logged a positive test result.

\subsection*{Reproducibility and algorithm immutability}

Reproducibility  depends on the ability to reproduce a given analysis from a version of the dataset and a set of algorithms run on the dataset. For this to be possible, algorithms must be considered immutable once implemented. This allows any subsequent version of the software to generate results consistent with those of the software version in which the algorithm was introduced.

ExeTera does this by requiring that a version of any given algorithm that is created is treated as immutable in the code base. This means that any target script is guaranteed to exhibit the same behaviour, provided that the following conditions hold. Firstly, any algorithms written for ExeTera are explicitly versioned. Secondly, any randomness introduced must be given consistent random seeds and, ideally, multiple sources of randomness should be given different random number generators. Once an algorithm is used in analysis, it may no longer be altered in the codebase, even if it subsequently shown to contain errors. This enables researchers to run multiple versions of the same algorithm as part of their analytics and understand how sensitive their results are to changes and corrections.
An example for this is the multiple versions of height / weight / body mass index (BMI) cleaning that have been devised over the course of the project; each is available as separate versions of the algorithm for reproducibility.

\subsection*{Implementation}

ExeTera is implemented in the Python programming language. Python has two aspects that make it suitable for writing software that performs data analytics and numerical analysis. Firstly, it is dynamically typed, which reduces code complexity and verbosity ~\cite{4273079}. Secondly, it has a strong ecosystem of scientific libraries and tools to mitigate the performance and memory penalties that come with using a dynamically typed, byte-code interpreted language and runtime.

\subsubsection*{Working with Python}

Code that is compiled and run directly in CPython (the reference Python implementation) executes in the Python interpreter. The Python interpreter is extremely slow relative to optimised code such as that generated by compiled, optimised C/C++; in many cases it is orders of magnitude slower. Python is dynamically typed, but its type system does not provide light-weight objects to represent primitive types. Even numeric values such as integers and floats are stored as full objects, and typically take up 28 bytes for a 4 byte integer value. This overhead precludes efficient memory usage when iterating over large numbers of values.

Numpy ~\cite{walt2011numpy, harris2020array} is the Python community’s main tool for circumventing such time and space inefficiencies. Amongst other features, it provides a library for space-efficient representations of multi-dimensional arrays, and a large library of time-efficient operations that can be carried out on arrays.

The performance of such operations can be orders of magnitude faster than native CPython, but this is conditional on minimising the number of transitions between Python code and the internal compiled code in which the operations are implemented.

Not all code can be easily phrased to avoid transitions between CPython and Numpy internals. Where this is not possible, Numba ~\cite{10.1145/2833157.2833162} is used to compile away the dynamic typing and object overhead, resulting in functions that execute at near optimised C performance levels.

\subsubsection*{Serialised data representations}

CSV is a very common way to portably represent large datasets, but it comes with many drawbacks, including  a lack of strong typing and an inability to rapidly index to a given location in the dataset. These issues become severely problematic at scale, and so an alternative serialised data representation is required.

\paragraph*{Data representation: row-local vs. column-local data formats}

Data storage formats can be classified as primarily row-local or primarily column-local. This choice has key implications for analytics software. Row-local data formats store groups of related fields for a given data entry together in memory. Column-local data formats store a specific value for a group of data entries together in memory.

\paragraph*{Row-local data format}

CSV format nearly always used in a row-local fashion, i.e. rows are data entries and columns are fields. This typically makes CSV very slow to parse for a subset of the data; with CSV this is exacerbated because escape-sequenced line-breaks mean that the entirety of each line must be parsed to determine the start of the next row. Even without this issue, row-local data formats suffer from locality of reference issues ~\cite{Drepper07whatevery}.

\paragraph*{Column-local data format}

Column-local data storage enables very efficient access to a given field. All the entries for a given field are (effectively) contiguous in memory and so loading a single field can be done in an asymptotically optimal fashion. When a dataset has many fields and a given operation operates on a smaller number of those fields, scaling the operation is a far simpler proposition, technically ~\cite{10.5555/1083592.1083658, stonebraker2010sql}.

The ability to load specific subsets of the data with maximal efficiency benefits both the ability to scale and the latency with which operations can be performed on a small subset of the dataset. As such, a column-local data format is a preferable format.

\paragraph*{HDF5 as the ExeTera serialised data format}

HDF5 ~\cite{hdf5} is a data format for storing keys and their associated values in a hierarchically organised, nested collection. In contrast with CSV, HDF5 stores data in a column-local format. It also allows for data to be stored as binary, concrete data types.
HDF5 permits a user to explore the overall structure of the data without loading fields. Fields are loaded at the point that a user specifically requests the contents of a given field. This can be a direct fetch of the entire field or an iterator over the field. This makes it a suitable initial data format for ExeTera, although alternative columnar data storage formats are being considered to replace HDF5 for future development due primarily to issues of dataset fragility and shortcomings relating to concurrent reading / writing.

\subsubsection*{Space-efficient operations}

Most analysis of tabular data is performed through a combination of joins, sorts, filters and aggregations. ExeTera operates on arrays of effectively unlimited length, particularly when certain preconditions are met, using the following techniques.

\subsubsection*{Sorting}

Sorting is one of the key operations that must scale in order to process large datasets, as imposition of a sorted order enables operations such as joins to scale. ExeTera uses several techniques to provide highly scalable sorting.

\paragraph*{Generation of a sorted index}

Rather than sorting data directly, ExeTera generates a sorted index that is a permutation of the original order. This is used to scale related sort operations, and implement a soft sort, where fields are stored in a natural sorted order and the permuted index applied when the field is read.

\paragraph*{Scaling multi-key sorts on long arrays}

Multi-key sorts are memory intensive when keys are large, and expensive due to the internal creation of tuples in the inner loops of sorts. Multi-key sorts in ExeTera are rephrased as a series of sorts on individual keys from right to left, where the output of each sort step is a sorted index that is the input to the next sort step, using a stable sort. Figure \ref{fig:multi_key_sort} shows pseudocode for this operation.

\paragraph*{Scaling sorts on very long arrays}

ExeTera has a second sorting algorithm that can be selected if an array is too large to fit into memory in its entirety. Such arrays are sorted via a two-phase approach in which the array is divided into subsets; each subset is sorted, and the sorted subsets are merged together by maintaining a heap of views onto the sorted subsets. A separate index is generated and maintained with the sorted chunks, so that the merge phase is stable. Figure \ref{fig:streaming_sort} shows pseudocode for this operation

\paragraph*{Sorting multiple fields}

The sorts described above, that produce a permutation of the original order, can be used to sort multiple fields in a space-efficient fashion. For large arrays, the array can be permuted in turn and written back to disk, or the permuted order maintained and reapplied when needed. ExeTera scales to provide this functionality even for very large arrays.

\paragraph*{Operations on sorted fields}

A number of operations become merges with various predicates when performed on fields that have been sorted by the key field and can be performed in $O(m+n)$ time where $m$ and $n$ are the lengths of the fields to be merged. This includes joins and aggregations. ExeTera performs these operations as merges when the key field is sorted. Importantly, arbitrarily large fields can be operated on in this way.

\subsubsection*{Joining}

\paragraph*{Generation of join maps}

Rather than performing the join on the fields themselves, ExeTera first generates primary key and foreign key index maps, which are then subsequently applied to the fields to be joined.

\paragraph*{Joining multiple fields}

As with sorts above, once the mapping indices have been calculated, they are applied to each field on the left side to map to the right side of the join, or vice-versa.

\paragraph*{Joining on sorted keys}

When the data is sorted on the keys of the respective fields, ExeTera rephrases joins as ordered merge operations.

\subsubsection*{Aggregation}

\paragraph*{Generation of aggregation maps / spans}

Aggregation is another operation that ExeTera optimises through use of pregenerated indices, particularly in the case that the data is sorted in aggregation key order.

\paragraph*{Aggregating on sorted keys}

As with joins, when the data is sorted on the keys of the aggregated fields, joins are performed by ExeTera in a very scalable and efficient fashion by precomputing spans representing ranges of the key field with the same key value. This can be iterated over, and aggregations performed in a streaming fashion.

\section*{Data availability}

The Covid Symptom Study dataset is available at \hyperlink{https://healthdatagateway.org}{https://healthdatagateway.org}, by searching for “COVID-19 Symptom Tracker Dataset”, or can be directly accessed at \hyperlink{https://web.www.healthdatagateway.org/dataset/fddcb382-3051-4394-8436-b92295f14259}{https://web.www.healthdatagateway.org/dataset/fddcb382-3051-4394-8436-b92295f14259}, at the time of writing.

Code for generating synthetic data is detailed in the \textit{Code availability} section.

\section*{Code availability} \label{codeavailability}

%The source code of ExeTera is made available through github at \hyperlink{https://github.com/KCL-BMEIS/ExeTera.git}{https://github.com/KCL-BMEIS/ExeTera.git} under the Apache 2.0 license, at the time of writing. ExeTera can be installed through pip and can be found at \hyperlink{https://pypi.org/project/exetera/}{https://pypi.org/project/exetera/}, at the time of writing. ExeTera Covid Symptom Study scripts for various analytics can be found at \hyperlink{https://github.com/KCL-BMEIS/ExeTeraCovid.git}{https://github.com/KCL-BMEIS/ExeTeraCovid.git}.

All source code for ExeTera is made available through github under the Apache 2.0 license, at the time of writing. The code is split up into three separate projects.

\paragraph{ExeTera}
The core functionality for ExeTera is hosted at \hyperlink{https://github.com/KCL-BMEIS/ExeTera.git}{https://github.com/KCL-BMEIS/ExeTera.git} and is available through pypi via \texttt{pip install exetera}.

\paragraph{ExeTeraCovid}
The functionality for the Covid Symptom Study dataset, (algorithms, scripts and other resources) is hosted at \hyperlink{https://github.com/KCL-BMEIS/ExeTeraCovid.git}{https://github.com/KCL-BMEIS/ExeTeraCovid.git} and is available through pypi via \texttt{pip install exeteracovid}. Installing \texttt{exeteracovid} installs \texttt{exetera}.

\paragraph{ExeTeraEval}
Code for creating evaluation datasets, including the dataset used to evaluate join operations, is hosted at \hyperlink{https://github.com/KCL-BMEIS/ExeTeraEval.git}{https://github.com/KCL-BMEIS/ExeTeraEval.git}.

\section*{Contribution statement}

BM, EK, MSG, and AV contributed to the software. MSG, CHS, EM, LSC, and MA provided feedback on the application programmer interface. RJ, JW and TS created the Covid Symptom Study that this software was created to provide data curation for. BM and MM wrote the manuscript. BM, EK, MSG, CHS, EM, LSC, MA, AV, ATC, PWF, RD, JW, TS, CJS, MM and SO reviewed and edited the draft. MM and SO supervised the project.

\bibliography{refs}

\section*{Figures \& Tables}

\begin{table}[ht]
\centering
\begin{tabular}{|l|l|l|l|l|}
\hline
& Patients & Assessments & Tests & Diet \\
\hline
Row count & 4,684,218 & 184,045,890 & 1,392,782 & 1,553,651 \\
\hline
Time to import (seconds) & 386.3 & 3849.6 & 20.1 & 52.9 \\
\hline
\end{tabular}
\caption{Import performance for the Covid Symptom Study snapshot dated 8th October, 2020.}
\label{tab:import}
\end{table}

\begin{table}[ht]
\centering
\begin{tabular}{| c | c | c | c | c | c | c |}
\hline
& \multicolumn{6}{c|}{Patient fields read} \\
\hline
Data source & N=1 & N=2 & N=4 & N=8 & N=16 & N=32 \\
\hline
Pandas read time (seconds) & 10.27 & 11.49 & 12.84 & 13.92 & 15.53 & 18.83 \\
\hline
Time to import (seconds) & 0.0242 & 0.0315 & 0.0465 & 0.0558 & 0.0787 & 0.0928 \\
\hline
\end{tabular}
\caption{\label{tab:read_patent_fields} Reading fields from the Covid Symptom Study 2020/10/08 snapshot, patient table. Times (in seconds) to load the first 1, 2, 4, 8, 16, and 32 fields respectively.}
\end{table}

\begin{table}[ht]
\centering
\begin{tabular}{| c | c | c | c | c | c | c |}
\hline
& \multicolumn{6}{c|}{Patient fields read} \\
\hline
Data source & N=1 & N=2 & N=4 & N=8 & N=16 & N=32 \\
\hline
Pandas read time (seconds) & 144.1 & X & X & X & X & X \\
\hline
Time to import (seconds) & 0.897 & 1.810 & 2.292 & 2.878 & 3.296 & 4.056 \\
\hline
\end{tabular}
\caption{\label{tab:read_assessment_fields} Reading fields from the Covid Symptom Study 2020/10/08 snapshot, assessment table. Times (in seconds) to load the first 1, 2, 4, 8, 16, and 32 fields respectively. X means that the operation could not be completed due to lack of memory.}
\end{table}

\begin{table}[ht]
\centering
\begin{tabular}{| c | c | c | c |}
\hline
left\_fk\_ids & right\_ids & right\_data\_0 & right\_data\_1 \\
\hline
0 & 0 & 51 & 36 \\
\hline
1 & 1 & 98 & 34 \\
\hline
1 & 2 & 31 & 47 \\
\hline
2 & 3 & 4 & 43 \\
\hline
4 & 4 & 49 & 18 \\
\hline
5 & 5 & 80 & 85 \\
\hline
5 & 6 & 43 & 20 \\
\hline
6 & 7 & 47 & 71 \\
\hline
8 & 8 & 97 & 87 \\
\hline
9 & 9 & 56 & 64 \\

\hline
\end{tabular}
\caption{\label{tab:join_data_example}The first ten rows of the dataset used to evaluate joins, in the two field case. The \texttt{right\_ids} field is the primary key for the right hand side and the \texttt{left\_fk\_ids} field is the foreign key on the left hand side. This data is generated by running the scripts provided in the ExeTeraEval github project, detailed in the code availability section.}
\end{table}

\begin{table}[ht]
\centering
\begin{tabular}{| c | c | c | c | c | c | c | c |}
\hline
\multicolumn{2}{|c|}{} & \multicolumn{6}{c|}{Time to left join fields (seconds)} \\
\hline
\multicolumn{2}{|c|}{Row count} & N=1 & N=2 & N=4 & N=8 & N=16 & N=32 \\
\hline
\multirow{2}{*}{1,048,576 (2$^{20}$)} & Pandas & \textbf{0.152} & \textbf{0.152} & \textbf{0.153} & \textbf{0.153} & \textbf{0.164} & \textbf{0.180} \\
                                    & ExeTera & 0.249 & 0.254 & 0.265 & 0.291 & 0.335 & 0.422 \\
\hline
\multirow{2}{*}{2,097,152 (2$^{21}$)} & Pandas & 0.281 & 0.283 & \textbf{0.285} & \textbf{0.291} &     \textbf{0.314} & \textbf{0.347} \\
                                    & ExeTera & \textbf{0.266} & \textbf{0.278} & 0.300 & 0.349 & 0.439 & 0.617 \\
\hline
\multirow{2}{*}{4,194,304 (2$^{22}$)} & Pandas & 0.710 & 0.721 & 0.727 & 0.741 & 0.773 & \textbf{0.837} \\
                                    & ExeTera & \textbf{0.297} & \textbf{0.322} & \textbf{0.369} & \textbf{0.460} & \textbf{0.646} & 1.02 \\
\hline
\multirow{2}{*}{8,388,608 (2$^{23}$)} & Pandas & 1.66 & 1.67 & 1.68 & 1.71 & 1.78 & 1.91 \\
                                    & ExeTera & \textbf{0.359} & \textbf{0.404} & \textbf{0.497} & \textbf{0.674} & \textbf{1.05} & \textbf{1.81} \\
\hline
\multirow{2}{*}{16,777,216 (2$^{24}$)} & Pandas & 3.96 & 3.96 & 3.99 & 4.06 & 4.18 & 4.44 \\
                                     & ExeTera & \textbf{0.474} & \textbf{0.563} & \textbf{0.746} & \textbf{1.09} & \textbf{1.81} & \textbf{3.27} \\
\hline
\multirow{2}{*}{33,554,432 (2$^{25}$)} & Pandas & 8.17 & 8.33 & 8.27 & 8.46 & 8.65 & 9.22 \\
                                     & ExeTera & \textbf{0.701} & \textbf{0.879} & \textbf{1.23} & \textbf{1.93} & \textbf{3.35} & \textbf{6.20} \\
\hline
\multirow{2}{*}{67,108,864 (2$^{26}$)} & Pandas & 18.9 & 19.1 & 19.0 & 19.2 & 19.8 & 20.8 \\
                                       & ExeTera & \textbf{1.17} & \textbf{1.52} & \textbf{2.21} & \textbf{3.62} & \textbf{6.40} & \textbf{12.2} \\
\hline
\multirow{2}{*}{134,217,728 (2$^{27}$)} & Pandas & 40.7 & 41.0 & 41.3 & 41.4 & 42.7 & \textit{X} \\
                                        & ExeTera & \textbf{2.09} & \textbf{2.80} & \textbf{4.23} & \textbf{6.96} & \textbf{12.6} & \textbf{23.7} \\
\hline
\multirow{2}{*}{268,435,456 (2$^{28}$)} & Pandas & \textit{X} & \textit{X} & \textit{X} & \textit{X} & \textit{X} & \textit{X} \\
                                        & ExeTera & \textbf{3.98} & \textbf{5.35} & \textbf{8.18} & \textbf{13.8} & \textbf{24.9} & \textbf{48.2} \\
\hline
\multirow{2}{*}{536,870,912 (2$^{29}$)} & Pandas & \textit{X} & \textit{X} & \textit{X} & \textit{X} & \textit{X} & \textit{X} \\
                                        & ExeTera & \textbf{7.72} & \textbf{10.4} & \textbf{16.2} & \textbf{26.9} & \textbf{50.3} & \textbf{95.9} \\
\hline
\multirow{2}{*}{1,073,741,824 (2$^{30}$)} & Pandas & \textit{X} & \textit{X} & \textit{X} & \textit{X} & \textit{X} & \textit{X} \\
                                        & ExeTera & \textbf{15.3} & \textbf{21.1} & \textbf{32.8} & \textbf{55.3} & \textbf{100.8} & \textbf{197.7} \\
\hline
\multirow{2}{*}{2,147,483,648 (2$^{31}$)} & Pandas & \textit{X} & \textit{X} & \textit{X} & \textit{X} & \textit{X} & \textit{X} \\
                                        & ExeTera & \textbf{30.2} & \textbf{42.8} & \textbf{65.6} & \textbf{112.1} & \textbf{241.8} & \textbf{602.3} \\
\hline
\end{tabular}
\caption{ Left join pandas \texttt{merge} vs. ExeTera \texttt{ordered\_merge\_left}: row count (rows) and fields joined count (columns). All times are in seconds. X means that the operation could not be completed due to lack of memory. Bold entries indicate the fastest join.}
\label{tab:left_join}
\end{table}

\begin{table}[ht]
\centering
\begin{tabular}{| c | c | c | c | c |}
\hline
& \multicolumn{4}{c|}{Journaling dataset snapshots} \\
\hline
Data source & Patients & Assessments & Tests & Diet \\
\hline
August 1st row count & 4,402,930 & 129,423,329 & 749,937 & 659 \\
\hline
September 1st row count & 4,480,270 & 153,655,115 & 991,128 & 1,291,237 \\
\hline
Rows only in old & 2,519 & 108,231 & 485 & 0 \\
\hline
Rows only in new & 86,301 & 243,40,017 & 241,676 & 1,290,578 \\
\hline
Rows updated & 1,632,849 & 702 & 18,169 & 630 \\
\hline
Rows not updated & 2,761,120 & 129,314,396 & 731,283 & 29 \\
\hline
Journaled row count & 6,122,080 & 153,764,048 & 1,009,782 & 1,291,867 \\
\hline
Time to import (seconds) & 145.1 & 2273 & 6.616 & 8.179 \\
\hline
\end{tabular}
\caption{\label{tab:journaling} Journaling Covid Symptom Study snapshots from 1st August 2020 and 1st September 2020. This table shows results in terms of row counts and the time taken to perform the journaling.}
\end{table}

\begin{figure}[ht]
\begin{lstlisting}
{
  "exetera": {
    "version": "1.0.0"
  },
  "schema": {
    "alpha": {
      "primary_keys": [ "a_pk" ],
      "fields": {
        "a_pk": { "field_type": "numeric", "value_type": "int32" },
        "field_x": { "field_type": "fixed_string", "length": 5 },
        "field_dt": { "field_type": "datetime" }
      }
    },
    "beta": {
      "primary_keys": [ "b_pk" ],
      "foreign_keys": {
        "a_fk": { "space": "alpha", "key": "id" }
      },
      "fields": {
        "b_pk": { "field_type": numeric, "value_type": "int64" },
        "a_fk": { "field_type": numeric, "value_type": "int32" },
        "field_m": {
          "field_type": "categorical",
          "categorical": {
            "value_type": "int8",
            "strings_to_values": { "": 0, "no": 1, "yes": 2 }
          }
        },
        "field_n": {
          "field_type": "categorical":,
          "categorical": {
            "value_type": "int8",
            "strings_to_values": { "left": 0, "right": 1 },
            "out_of_range": "freetext"
          }
        },
        "field_dt": {"field_type": "datetime", "optional": "True"}
      }
    }
  }
}
\end{lstlisting}
\caption{An illustrative, minimal example of an ExeTera JSON schema for use when importing data from CSV.}
\label{fig:schema_file_example}
\end{figure}

\begin{figure}[ht]
\begin{lstlisting}
    F: a list of unsorted data fields that are being sorted on
    Du: the unsorted data for a given field
    Ds: the sorted data for a given field
    I: Index from sorting data with an 'argsort' function  
    A: Accumulated sorted index
    getdata: a function that fetches the data from a field
    range: a function that returns integer values between start (inclusive)
           and end (exclusive)
    argsort: a function that returns the permutation of indices by sorting a
             vector of data
    permute: a function that takes a data field and an index and applies the
             latter to the former 
    
    Du = getdata(F[0])
    Ds = permute(Du, A)
    I = argsort(Ds)
    A = permute(A, I)
    for i in range(1, n):
        Du = getdata(F[i])
        Ds = permute(Du, A)
        I = argsort(Ds)
        A = permute(A, I)
    return A_n
\end{lstlisting}
\caption{\label{fig:multi_key_sort}Pseudocode for a multi-key sort that outputs a sorted index for subsequent application to many fields.}

\end{figure}

\begin{figure}[ht]
\begin{lstlisting}
    D: the data vector to be sorted
    Ds: the sth  subset of D
    I: a set of unpermuted indices (the indices of each element of D)
    Is: the sth subset of I
    subsets: a function that splits a very large sequence up into smaller subsets
    sort_and_apply: a function that sorts a data vector and permutes an index
                    vector accordingly 
    cursor: a view onto a vector that amortises the cost of reading individual
            elements of that vector from a drive
    best: a function that selects the next element from a collection of data and
          index cursors
    next: a function that increments a cursor
    
    Ic = list()
    Dc = list()
    For s in subsets(D):
        Dis, Dps  = sort_and_apply(Ds, Is)
        Ic.append(cursor(Dis))
        Dc.append(cursor(Dps))
    Dr = []
    For i in length(D):
        j = best(Ic, Dc)
        Dr.append(Dc[j])
        next(Ic[j])
        next(Dc[j])
\end{lstlisting}
\caption{\label{fig:streaming_sort}Pseudocode for a streaming sort that outputs a sorted index for subsequent application to many fields.}

\end{figure}

\end{document}